\documentclass[conference]{IEEEtran}
\IEEEoverridecommandlockouts
\usepackage{cite}
\usepackage{amsmath,amssymb,amsfonts,multirow}
\usepackage{graphicx}
\usepackage[normalem]{ulem}
\usepackage[table,xcdraw]{xcolor}
\usepackage{textcomp}
\usepackage{xcolor}
\usepackage{ctable}
\usepackage{subcaption}
\usepackage{soul}
\usepackage{multicol}
\usepackage[ruled,vlined,linesnumbered]{algorithm2e}
\usepackage{algpseudocode}
\usepackage[bindingoffset=0.05in,
            top=0.75in, 
            bottom=1in, 
            left=0.625in, 
            right=0.625in]{geometry}

\usepackage{authblk}
 
\def\BibTeX{{\rm B\kern-.05em{\sc i\kern-.025em b}\kern-.08em
 T\kern-.1667em\lower.7ex\hbox{E}\kern-.125emX}}
\begin{document}

\newcommand{\CLASSINPUTtoptextmargin}{0.75in}
\newcommand{\CLASSINPUTbottomtextmargin}{1.01in}
\setlength{\columnsep}{0.24in} 

\title{GreenFLag: A Green Agentic Approach for Energy-Efficient Federated Learning}

\author{Theodora Panagea*, Nikolaos Koursioumpas*, Lina Magoula*, Ramin Khalili**
\\
* \emph{Dept. of Informatics and Telecommunications, National and Kapodistrian University of Athens, Greece} 
\\
** \emph{Huawei Heisenberg Research Center (Munich), Germany}
\\

\{dpanagea*, nkoursioubas*, lina-magoula*\}@di.uoa.gr \\

\{ramin.khalili**\}@huawei.com}

\maketitle

\begin{abstract}
Progressing toward a new generation of mobile networks, 
a clear focus on integrating distributed intelligence across the system is observed to drive performance, autonomy, and real-time adaptability.
Federated learning (FL) stands out as a key emerging technique, enabling on-device model training while preserving data locality. However, its operation introduces substantial energy and resource demands. Energy needs are mostly met by grid power sources, while FL resource orchestration strategies remain limited. This work introduces GreenFLag, an agentic resource orchestration framework designed to minimize the energy consumption from the grid power to complete FL workflows, guarantee FL model performance, and reduce grid power reliance by incorporating renewable sources into the system. GreenFLag leverages a Soft-Actor Critic reinforcement learning approach to jointly optimize computational and communication resources, while accounting for communication contention and the dynamic availability of renewable energy. Evaluations using a real-world open dataset from Copernicus, demonstrate that GreenFLag significantly reduces grid energy consumption by $94.8\%$ on average, compared to three state-of-the-art baselines, while primarily relying on green power.

\end{abstract}

\section{Introduction}\label{Intro}
The rapid expansion of wireless networks and edge computing has significantly increased energy demands in the Information and Communication Technology (ICT) sector. While the sector currently accounts for roughly $4\%$ of global electricity consumption, projections indicate a rise to $10$–$20\%$ by 2030 \cite{LANGE2020106760}. At the same time, innovative, AI-driven concepts are emerging to enhance network automation and support new and challenging use cases and applications. One such concept is Federated Learning (FL), a decentralized Artificial Intelligence (AI) approach that enables cooperative model training among multiple devices without transferring any raw data. Although FL improves data privacy and reduces resilience on centralized data centers, its environmental impact has become a significant concern. As networks progress toward Beyond-5G and 6G with billions of connected devices, enhancing the energy efficiency of FL will be essential for long-term sustainability. Energy-efficient FL could reduce grid reliance yet it requires precise management of grid energy consumption, as it has a negative environmental impact through its associated $\mathrm{CO_2}$ emissions.

Integrating renewable energy into the network ecosystem reduces dependence on the grid. Energy harvesting technologies, including solar panels and wind turbines, can supply devices with locally generated power. However, fluctuations in these sources lead to uneven  energy reserves across devices due to weather patterns, geographic location, and hardware capabilities. This variability changes how much energy a device can spend on computation or communication at any given moment, imposing new constraints on resource-allocation strategies. In FL systems, where training and uplink transmissions demand substantial energy, effective operation requires intelligent scheduling mechanisms that can manage task execution with \textit{green} energy availability as a driver, while minimizing reliance on the grid. 


To address the growing energy and carbon footprint of distributed intelligence, global standard bodies have formalized energy efficiency frameworks and sustainable architecture principles. 
3GPP introduced studies for energy-saving management across 5G networks, and proposed advanced New Radio (NR) -level savings, treating energy efficiency as a core service requirement\cite{3gpp_ts28310}\cite{3gpp_tr38864}\cite{3gpp_tr22882}. 
ITU-R  provided methodologies for assessing mobile network efficiency and integrating renewable sources into ICT infrastructures\cite{itu_l1331}\cite{itu_l1383}\cite{itu_lsupp43}.
ETSI complements the above by standardizing energy KPIs and hybrid power solutions for 5G sites\cite{etsi_es203228}\cite{etsi_es203700}. 
IEEE has published technologies to evaluate and optimize energy efficiency at the architecture and processing level\cite{ieee_19231}\cite{ieee_19241}. 

Researchers have explored various techniques to improve efficiency. In \cite{WANG9488756} \cite{CHEN9712615} \cite{akhtarshenas2024federated}, the authors propose energy-aware FL solutions to meet latency goals and optimize bandwidth allocation. However, they overlook renewable integration or resource orchestration.  In \cite{li2024fedcarbon}, Li et al. propose FedCarbon, a carbon-efficient framework that includes client sampling and model pruning, in order to align training with periods of low carbon intensity.
Beyond FL-specific efforts, energy-aware orchestration has been explored for edge workloads. The authors in \cite{kaur2019multi}\cite{liao2024rethinking}\cite{patel2023advancements} explore sustainable power provisioning for edge/cloud. 
Reinforcement Learning (RL) has emerged as a powerful tool for dynamic resource allocation in energy-aware networks. 
By learning adaptive strategies, RL agents can optimize long-term goals like energy efficiency. 
Recent applications include managing transmission policies in energy-harvesting networks and optimizing FL processes\cite{wang2023reinforcement}\cite{XIA10.1145/3676536.3676815}. In \cite{feng2024joint}, the authors propose a joint scheduling and quantization optimization scheme for energy harvesting-enabled FL networks, so as to minimize training loss under strict energy causality constraints. In \cite{alvi2026constrained}, aconstrained Soft Actor-Critic approach is introduced to optimize joint computation offloading and resource allocation while handling long-term latency and energy constraints. 
Wiesner et al \cite{WIESNER10.1145/3632775.3639589} present an FL system that operates solely on excess renewable energy and spare computational capacity to reduce its carbon emissions to zero. However, they do not take into account the communication part of a network. In \cite{liao2025greenfl}, Liao et al investigate the carbon-efficient exploitation of spatio-temporal renewable energy variations across distributed edge computing systems.
Previous work demonstrated a significant reduction in FL energy consumption by adjusting device CPU frequencies and transmission powers. However, this approach treated all energy sources uniformly, ignoring opportunities to prioritize renewables. 

Overall, while FL research increasingly addresses communication and energy constraints, it generally lacks integration with renewable energy considerations. Conversely, sustainability frameworks rarely address FL-specific performance trade-offs. GreenFLag addresses this gap by jointly optimizing FL efficiency and renewable energy availability. 
To our knowledge, this is one of the few works to embed renewable awareness directly into the FL resource-orchestration loop. 

In this paper, we extend prior work \cite{koursioumpas2024safe} by explicitly incorporating renewable sources into the network system to further reduce the carbon footprint associated with grid power. GreenFLag adopts an agentic resource orchestration paradigm, in which an autonomous learning agent dynamically allocates computation and communication resources across FL iterations to minimize long-term grid energy consumption while preserving FL performance. The proposed RL-based agent not only controls the computation and communication resources of each device but also arranges FL tasks so that renewable energy becomes priority, with grid power serving only as a fallback.

To ensure practicality, we introduce a bandwidth scheduler that works alongside the RL agent. While the agent allocates bandwidth, the scheduler regulates competing transmissions to prevent channel congestion, preserving feasibility under realistic network conditions. 

\begin{table}[ht!]
\centering
\resizebox{\columnwidth}{!}{%
\begin{tabular}{|l|c|c|c|c|}
\hline
\textbf{Methodology Group} & \textbf{\begin{tabular}[c]{@{}c@{}}Agentic \end{tabular}} & \textbf{\begin{tabular}[c]{@{}c@{}}Optimization \end{tabular}} & \textbf{\begin{tabular}[c]{@{}c@{}}Green \end{tabular}} & \textbf{\begin{tabular}[c]{@{}c@{}}Network\end{tabular}} \\ \hline
\begin{tabular}[c]{@{}l@{}}1: Autonomous Orchestration \\ (\cite{wang2023reinforcement}, \cite{XIA10.1145/3676536.3676815}, \cite{alvi2026constrained}, \cite{koursioumpas2024safe})\end{tabular} & \checkmark & \checkmark & \checkmark &  \\ \hline
\begin{tabular}[c]{@{}l@{}}2: Spatio-Temporal Scheduling \\ (\cite{li2024fedcarbon}, \cite{WIESNER10.1145/3632775.3639589})\end{tabular} &  & \checkmark & \checkmark &  \\ \hline
\begin{tabular}[c]{@{}l@{}}3: Cross-Layer Provisioning \\ (\cite{WANG9488756, CHEN9712615, akhtarshenas2024federated}, \cite{kaur2019multi, liao2024rethinking, patel2023advancements}, \cite{feng2024joint})\end{tabular} &  & \checkmark &  &  \\ \hline
\textbf{GreenFLag (This Work)} & \textbf{\checkmark} & \textbf{\checkmark} & \textbf{\checkmark} & \textbf{\checkmark} \\ \hline
\end{tabular}%
}
\caption{Technical Comparison}
\label{sotatable}
\end{table}

 As illustrated in Table \ref{sotatable}, GreenFLag represents an advancement over current methodologies. In this table, \textit{Joint. Opt.} refers to the joint optimization of computational and communication resources, \textit{Green Aware} denotes the integration of renewable energy or sustainability metrics, \textit{Network} indicates the system's awareness of network contention and shared-channel interference. 
 
 While \textit{Group 1} leverages the flexibility of RL to handle stochastic environments, these works often assume best-effort networks. \textit{Group 2} and \textit{Group 3} frameworks provide insights into grid dynamics and hardware constraints respectively, but lack the autonomy for long-term optimization. GreenFLag addresses these limitations by including network contention awareness into its agentic orchestration, ensuring feasible resource allocation decisions under realistic network conditions, while also significantly reducing grid reliance.

The key contributions of this work can be summarized as follows: 
\begin{enumerate}
    \item A system model and problem formulation jointly optimizing computation and communication resources by integrating renewable energy into the system design. 
    \item A green energy-first strategy that prioritizes green energy consumption over grid power.
    \item A reinforcement learning-based agent that jointly orchestrates the computation capacity, the transmission power and the allocated bandwidth for each device.
    \item A penalty-based safety mechanism that enforces performance targets, while enabling adaptive resource allocation. 
    \item Introduction of a  scheduler, ensuring realistic bandwidth allocation and preventing over-provisioning in shared communication channels.  
    \item  A realistic evaluation framework using real-world renewable energy data from Copernicus to demonstrate that GreenFLag achieves on average $94.8\%$ reduction in carbon footprint compared to state-of-the-art baseline approaches, without compromising FL accuracy or convergence speed.  
\end{enumerate}

The rest of the paper is organized as follows. Section
II provides the system model. Section III provides the problem
formulation. Section IV describes the proposed solution that is
evaluated in Section VI using the simulation setup of Section
V. Finally, section VII concludes the paper.

\section{System Model} \label{system_model}
We consider an AI-enabled wireless network integrating a Federated Learning (FL) process, consisting of one central FL coordinator and $\mathcal{K}$ distributed edge devices acting as workers. All workers cooperatively contribute to the training of a shared Neural Network (NN). 
\begin{figure}[ht]
    \includegraphics[width=1\linewidth]{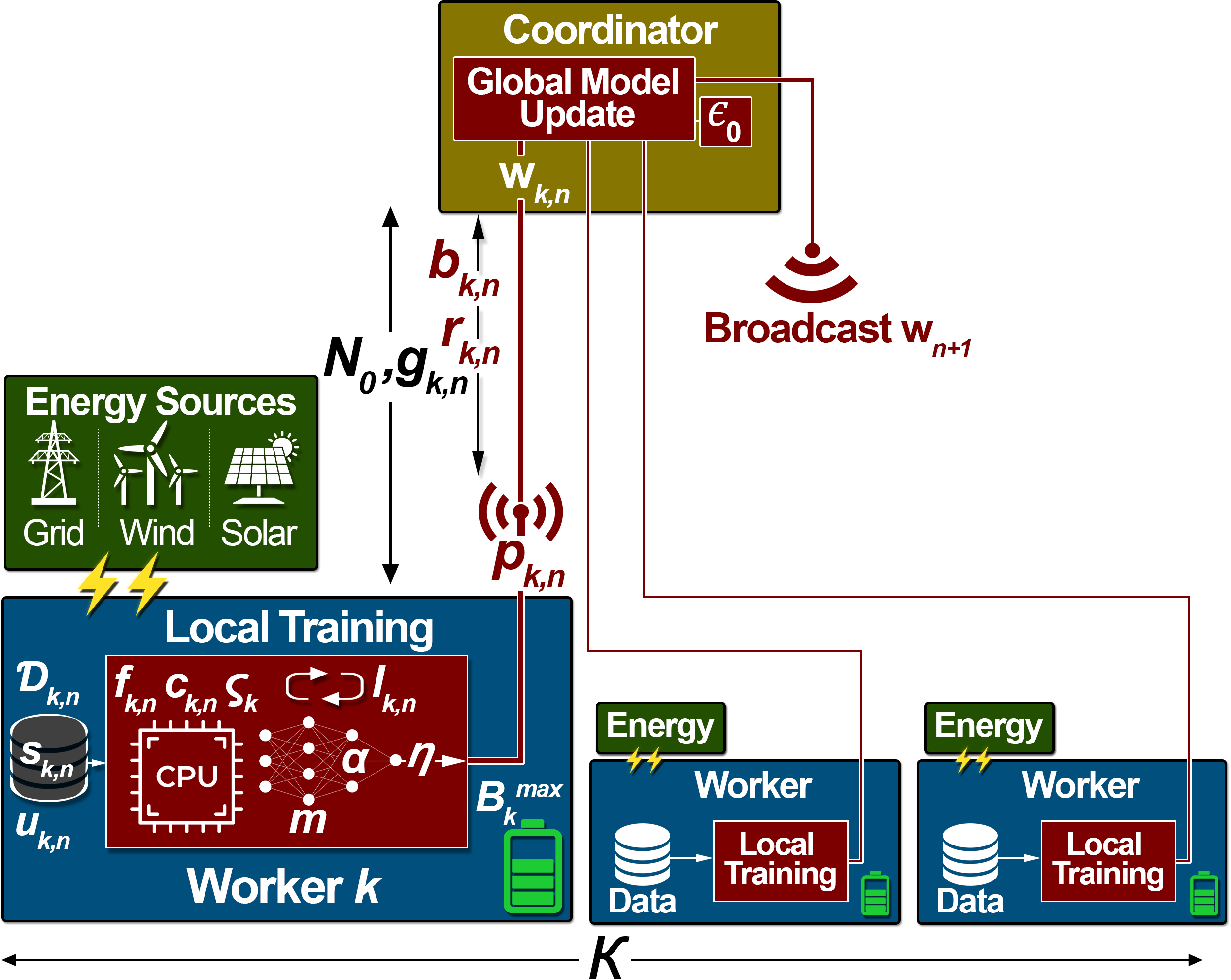}
    \caption{System Model}
    \label{fig_sysmodel}
    \vspace{-5pt}
\end{figure}

An FL process is comprised of a number of \textit{global} iterations, each one denoted by $n \in \mathbb{N}$. Each iteration $n$ contains three main phases. \\
\textbf{First Phase [Model Sharing]}: The FL coordinator distributes the shared model parameters $\textbf{w}_n \in \mathbb{R}$ of size $m \in \mathbb{R}_+$ in bits and complexity $\alpha \in \mathbb{R}_+$ in Floating Point Operations (FLOPs) to the involved workers\footnote{It is assumed that the coordinator energy consumption to broadcast the model is constant and negligible compared to the subsequent phases of the FL.}.
\\
\textbf{Second Phase [Computation]}: Each worker $k \in \mathcal{K}$ after receiving $\textbf{w}_n$, performs a local training using its own dataset $\mathcal{D}_{k,n}$ of $s_{k,n} \in \mathbb{N}$ samples, requiring $I_{k,n} \in \mathbb{N}$ \textit{local} iterations to reach a pre-selected performance target $\eta \in [0,1]$. Based on the  computational capacity $f_{k,n}$ (CPU speed) and its effective switched capacitance $\varsigma_k \in \mathbb{R}_+$, the worker $k$ can complete a certain number of FLOPs per cycle $c_{k,n} \in \mathbb{R}_+$. By $\tau_{k,n} \in \mathbb{R}_+$ and $E_{k,n}^{C} \in \mathbb{R}_+$ we denote the time and computation energy required by worker $k$ to complete a local training process. \\
\textbf{Third Phase [Communication]}: At the end of a local training, each worker transmits its updated model parameters $\textbf{w}_{k,n}$ to the FL coordinator for aggregation. The communication channel is modeled as a flat-fading with Gaussian noise power density $N_0 \in \mathbb{R}$ and channel gain $g_{k,n} \in \mathbb{R}$, where the fading is assumed constant. Let $b_{k,n} \in \mathbb{R}_+$ and $p_{k,n} \in \mathbb{R}_+$ be the assigned bandwidth and transmission power to worker $k$ at iteration $n$, respectively.
By $r_{k,n} \in \mathbb{R}_+$, $t_{k,n} \in \mathbb{R}_+$ and $E_{k,n}^{T} \in \mathbb{R}_+$ we denote the achievable data rate, the required time and communication energy to upload $\textbf{w}_{k,n}$ to the coordinator, respectively. The FL is realized in a synchronized manner. A \textit{global} iteration is finished when the coordinator receives updates from all workers or when a pre-selected time threshold $\mathsf{H} \in \mathbb{R}_+$ is reached. All workers should transmit their updates within $\mathsf{H}$. Updates from workers who have not met the time threshold are considered invalid and are not used to update the global model, and thus their used energy is deemed as wasted denoted by $E_{k,n}^W$. The coordinator produces an updated model $\textbf{w}_{n+1}$ from the received updates, and broadcasts it to the workers before they start the next \textit{global} iteration ($n+1$). 

These three phases are repeated until convergence to a pre-selected performance target $\epsilon_0 \in [0,1]$.\\
\textbf{Renewable energy integration:}
Each worker has access to energy harvested from renewable sources. The most widely used sources are considered, namely the solar and wind. 
The renewable energy harvested during $\mathsf{H}$, where a worker $k$ has access to at the $n^{th}$ \textit{global} iteration is formulated as:
\begin{equation} \label{renewable_energy_formulation}
 E^{R}_{k,n} = E^{S}_{k,n} + E^{WN}_{k,n}, 
\end{equation}
where $E^{S}_{k,n}$ and $E^{WN}_{k,n}$ denote the solar and wind energy harvested, respectively.\\
\underline{Solar energy}:
\begin{equation} \label{solar_energy}
E^{S}_{k,n} = P^{SE}_{k,n} \cdot \mathsf{H} \cdot \mathsf{z} ,
\end{equation}
where $P^{SE}_{k,n}$ is the effective solar radiation and $\mathsf{z}$ denotes the solar panel area. The $P^{SE}_{k,n}$ is provided by:
\begin{equation}
  P^{SE}_{k,n} = CL_n \cdot P^{SA}_{n},
\end{equation}
where $P^{SA}_{n}$ denotes the solar radiation of the atmosphere and the clearness index $CL_n \in [0.25,1]$ is given by: 
\begin{equation}
  CL_n = 1 - \frac{3}{4} \left( \frac{N_n}{8} \right) ^{3.4},
\end{equation}
where $N_n$ is the cloud cover expressed in oktas (0–8 scale).\\
\underline{Wind energy}:
\begin{equation} \label{wind}
  E^{WN}_{k,n} = P^{WN}_{n} \cdot \mathsf{H}, 
\end{equation}
where $P^{WN}_{n}$ is the average wind power density, defined as:
\begin{equation}
  P^{WN}_{n} = \int_{\mathsf{v}_{i,n}}^{\mathsf{v}_{f,n}} P(\mathsf{v})\cdot f(\mathsf{v})d\mathsf{v} ,
\end{equation}
which weights the instantaneous wind-power function
\begin{equation} \label{wind_power}
  P(\mathsf{v})= \frac{1}{2} \cdot \rho \cdot S \cdot \mathsf{v}^{3} 
\end{equation}
by the Weibull probability density function $f(\mathsf{v})$ that captures local wind conditions. The $\mathsf{v}_{i,n}, \mathsf{v}_{f,n}$ denote the measured wind speeds at the beginning and at the end of $\mathsf{H}$, $\rho$ is the air density, $S$ denotes the turbine sweep area, and $\mathsf{v}$ is the wind speed.
\\
\textbf{Battery Storage}: Each worker is equipped with a battery with maximum capacity $B^{\max}_{k}$. At each \textit{global} iteration $n$, each worker $k$ involved in the FL process incurs a total energy demand: $E_{k,n}^{\text{total}} = E_{k,n}^{C}+E_{k,n}^{T}$, corresponding to computation and communication FL tasks. This demand is satisfied by following a hierarchical energy usage order: the worker first consumes the renewable energy harvested during the iteration, denoted by $E^{R}_{k,n}$. If $E^{R}_{k,n}$ is insufficient to cover $E^{\text{total}}_{k,n}$, the worker then draws energy from its battery, provided that stored energy is available. In case the combined renewable supply and battery storage still fall short, 
the worker relies on grid 
energy $E^{G}_{k,n}$ to meet the deficit. If the harvested  energy exceeds the required consumption, the surplus is stored in the battery as $E^{B}_{k,n}$. 
The energy demands of each worker at each iteration are covered in the following order: (1) harvested energy $E_{k,n}^{R}$, (2) battery stored renewable energy $E_{k,n}^{B}$, and (3) grid energy $E_{k,n}^{G}$, in cases where demand exceeding renewable availability. Table \ref{table_notations} summarizes the notations.
\begin{table}[ht!]
 \renewcommand{\arraystretch}{1.05}
 \begin{tabular}{|p{0.048\textwidth}|p{0.39\textwidth}|}
\hline
\textbf{Params} & \textbf{Description} \\
\hline
$\mathcal{K}$ & Set of workers \\ \hline
$n$ & Index of the \textit{global} iteration (FL round) \\ \hline
$N_0$ & White Gaussian noise power spectral density\\ \hline
$g_{k,n}$ & Gain of the wireless channel the worker $k$ has access to at the $n^{th}$ \textit{global} iteration\\ \hline
$f_{k,n}$ & Available computational capacity of worker $k$ at the $n^{th}$ \textit{global} iteration\\ \hline
$b_{k,n}$ & Bandwidth assigned to worker $k$ at the $n^{th}$ \textit{global} iteration\\ \hline
$p_{k,n}$ & Transmission power of worker $k$ at the $n^{th}$ \textit{global} iteration \\ \hline
$r_{k,n}$ & Achievable transmission data rate of worker $k$ at the $n^{th}$ \textit{global} iteration\\ \hline
$\mathcal{D}_{k,n}$ & Local dataset of worker $k$ at the $n^{th}$ \textit{global} iteration\\ \hline
$\mathcal{L}_{k,n}$ & The set of groundtruth data of worker $k$ at the $n^{th}$ \textit{global} iteration\\ 
\hline
$s_{k,n}$ & Total number of data samples of worker $k$ at the $n^{th}$ \textit{global} iteration\\ \hline
$v_{k,n}$ & Dataset variance of worker $k$ at the $n^{th}$ \textit{global} iteration\\ \hline
$c_{k,n}$ & Total number of Floating Point Operations (FLOPs) per cycle that the worker $k$ can complete at the $n^{th}$ \textit{global} iteration\\ \hline
$\varsigma_k$ & Effective switched capacitance of worker $k$ \\ \hline
$\textbf{w}_n$ & Global FL model produced at the $n^{th}$ \textit{global} iteration \\ \hline
$\alpha$ & Complexity of the global FL model in terms of total number of Floating Point Operations (FLOPs) \\ \hline
$m$ & Size of global FL model in bits \\ \hline
$I_{k,n}$ & Number of \textit{local} iterations required to reach $\eta$ at the worker $k$ at the $n^{th}$ \textit{global} iteration\\ \hline
$\tau_{k,n}$ & The time required by a worker $k$ to complete a local training process at the $n^{th}$ \textit{global} iteration\\ \hline 
$t_{k,n}$ & The time required by worker $k$ to transmit its model updates at the $n^{th}$ \textit{global} iteration\\ \hline
$\textbf{w}_{k,n}$ & Model parameters of worker $k$ at the $n^{th}$ \textit{global} iteration \\ \hline
$\eta$ & Training performance target of all workers\\ \hline
$\epsilon_0$ & Training performance target of the global FL model\\ \hline
$E^C_{k,n}$ & Computation energy of worker $k$ at the $n^{th}$ \textit{global} iteration\\ \hline
$E^T_{k,n}$ & Transmission energy of worker $k$ at the $n^{th}$ \textit{global} iteration\\ \hline
$\mathsf{H}$ & Pre-selected time threshold  \\ \hline
$E^W_{k,n}$ & Wasted energy of worker $k$ at the $n^{th}$ \textit{global} iteration\\ \hline
$E^R_{k,n}$ & Total harvested energy of worker $k$ at the $n^{th}$ \textit{global} iteration\\ \hline
$E^S_{k,n}$ & Harvested solar energy of worker $k$ at the $n^{th}$ \textit{global} iteration\\ \hline
$E^{WN}_{k,n}$ & Harvested wind energy of worker $k$ at the $n^{th}$ \textit{global} iteration\\ \hline
$P^{SE}_{k,n}$ & Effective solar radiation of worker $k$ at the $n^{th}$ \textit{global} iteration\\ \hline
$P^{SA}_{n}$ & Direct Solar radiation at the $n^{th}$ \textit{global} iteration\\ \hline
$CL_{n}$ & Clearness Index at the $n^{th}$ \textit{global} iteration\\ \hline
$N_{n}$ & Cloud cover at the $n^{th}$ \textit{global} iteration\\ \hline
$P^{WN}_{n}$ & Average wind power density at the $n^{th}$ \textit{global} iteration\\ \hline
$\mathsf{v}$ / $\rho$ & Wind speed / Air density\\ \hline
$\mathsf{z}$ / $S$ & Solar panel area / Turbine sweep area\\ \hline
$E^W_{k,n}$ & Wasted energy of worker $k$ at the $n^{th}$ \textit{global} iteration\\ \hline
$E^{total}_{k,n}$ &Total consumed energy of worker $k$ at the $n^{th}$ \textit{global} iteration\\ \hline
$E^B_{k,n}$ & Surplus harvested energy of worker $k$ at the $n^{th}$ \textit{global} iteration\\ \hline
$B^{\max}_{k}$ & Maximum capacity of worker's $k$ local battery\\ \hline
$E^G_{k,n}$ & Grid consumed energy of worker $k$ at the $n^{th}$ \textit{global} iteration\\ \hline

\end{tabular}
\caption{Notation Table} 
\label{table_notations} 
\end{table}



\section{Problem Formulation} \label{problem_formulation}
 The objective of our problem formulation is to achieve energy efficiency by minimizing the carbon footprint of  the system, i.e the minimization of the workers' overall energy consumption from the grid power $E^{G}_{total}$, while guarantying a certain FL model performance target $\epsilon_0$. Our optimization problem is  modeled  as  a  Markov Decision Process (MDP) \cite{Wiley1994}, where the objective is to minimize the total grid energy consumption. 
 The objective function can be defined as:
\begin{equation} \label{objective_function}
\begin{aligned}
   \min_{ \textbf{f}_n, \textbf{p}_n, \textbf{b}_n} E^{G}_{total} 
   &= \sum_{n=1}^{I_0} \gamma^{n-1}\cdot\sum_{k=1}^{\mathcal{K}}E_{k,n}^{G}\\
   &= \sum_{n=1}^{I_0} \gamma^{n-1}\cdot\sum_{k=1}^{\mathcal{K}} \Omega_{k,n} \cdot \big( E_{k,n}^C + \\
   &+ E_{k,n}^T - (E_{k,n}^R + E_{k,n}^B)\big)_+,
\end{aligned}
\end{equation}
where $I_0$ denotes the terminal state, i.e. the \textit{global} iteration in which the FL model reaches the pre-selected performance target $\epsilon_0$, $\gamma \in [0,1]$ is the discount rate to account for the relative importance of the energy consumption of future \textit{global} iterations. The $\textbf{f}_n =[f_{1,n},...,f_{\mathcal{K},n}]^T$, $\textbf{p}_n = [p_{1,n},...,p_{\mathcal{K},n}]^T$, and $\textbf{b}_n = [b_{1,n},...,b_{\mathcal{K},n}]^T$ represent the computational capacity, the transmission power and the bandwidth of all workers to be optimized at each \textit{global} iteration $n$ of the FL process. 

The computation and communication energies ($E_{k,n}^C, E_{k,n}^T$) consumed at each iteration $n$ by worker $k$ are given by:
\begin{equation}  \label{computation_energy}
  E_{k,n}^C = \frac{\varsigma_{k} \cdot I_{k,n} \cdot \alpha \cdot s_{k,n} \cdot f_{k,n}^2} {c_{k,n}},
\end{equation}
and
\begin{equation} \label{transmission_energy}
  E_{k,n}^T = \frac{m \cdot p_{k,n}}{r_{k,n}},
\end{equation}
where
\begin{equation}
  r_{k,n} = {b_{k,n} \cdot log_2\left(1 + \frac{g_{k,n} \cdot p_{k,n}}{b_{k,n} \cdot N_0}\right)}.
\end{equation}

The $(\cdot)_+$ discards negative values and $\Omega_{k,n}$ is an indicator function defined as:
\begin{equation} \label{indicator_function_omega}
\begin{aligned}
 \Omega_{k,n} = 
 \left\{
 \begin{array}{ll}
 1, & p_{k,n} > 0 \text{ and } b_{k,n} > 0 \text{ and } f_{k,n} > 0\\
 0, & \text{otherwise}
 \end{array}
 \right.
 \end{aligned}
\end{equation}

The $E^B_{k,n}$ denotes the amount of renewable energy stored at the local battery of worker $k$, during \textit{global} iteration $n$. 
The battery energy $E^B_{k,n}$ is used only when the harvested renewable energy $E^{R}_{k,n}$ is insufficient to cover $E^{\text{total}}_{k,n}$. The battery state evolves according to: 
\begin{equation} \label{battery_contr_next}
  E_{k,n}^B = \min{ \{ B_k^{\max}, \big( E^B_{k,n-1} + (E^R_{k,n} - E^{\text{total}}_{k,n}) \big)_+\} },
\end{equation}
so that any surplus renewable energy $(E^R_{k,n} > E^{\text{total}}_{k,n})$ is stored in the battery up to its maximum capacity. 

In order to efficiently enforce the actions related to the bandwidth allocation ($\textbf{b}_n$), a First Come First Served (FCFS) bandwidth scheduler is introduced as part of the system. More specifically, given a per worker resource assignment ($f_{k,n},p_{k,n}, b_{k,n}$):

\begin{enumerate}
    \item After completing its computation tasks, which take duration $\tau_{k,n}$, worker $k$ attempts to access the communication channel and requests its assigned bandwidth $ b_{k,n}$
    to transmit its model updates.
    \item The FCFS bandwidth scheduler then checks whether the channel has sufficient remaining capacity to allocate the requested bandwidth for worker $k$ \label{step_3}.
    \begin{itemize}
    \item In case of bandwidth availability, the worker accesses the channel and proceeds with the transmission task. When the worker completes its transmission task, it releases its allocated bandwidth. 
    \item If there is no sufficient bandwidth to allocate, the worker $k$ joins a queue. The scheduler repeatedly checks bandwidth availability until either enough capacity becomes available or the time threshold $\mathsf{H}$ is reached. The time that worker 
    $k$ spends waiting in the queue is denoted by
    $tq_{k,n}$. 
\end{itemize}
\end{enumerate}

The objective function should be subject to a number of constraints to ensure feasibility of assigned resources. The complete list of constraints is provided below.
\begin{equation} 
\label{complete_version_constraint_total_latency}
\begin{aligned}
  \tau_{k,n} + tq_{k,n} + tr_{k,n} < \mathsf{H}, \forall k \in \mathcal{K} 
  \end{aligned}
\end{equation}

\begin{equation} \label{complete_version_constraint_per_device_capacity}
\begin{aligned}
 0\leq f_{k,n} \leq f_{k,n}^{max}, \forall k \in \mathcal{K} 
\end{aligned}
\end{equation}

\begin{equation} 
\begin{aligned}
\label{complete_version_constraint_per_device_power}
 0 \leq p_{k,n} \leq p_{k,n}^{max} , \forall k \in \mathcal{K}     
 \end{aligned}
\end{equation}

\begin{equation} 
\label{complete_version_constraint_idle}
\begin{aligned}
 \sum_{k=1}^{\mathcal{K}} f_{k,n} > 0,
\end{aligned}
\end{equation}
\begin{equation}
\label{complete_version_constraint_total_bandwidth}
\begin{aligned}
  0 \leq b_{k,n} &\leq b^{\text{max}}_n - \sum_{i \in \mathcal{K}, i\neq k} \left( b_{i,n} \cdot ch^{(q)}_{k,n} \right),\\  &\forall k \in \mathcal{K}, \forall q \in \left[\tau_{k,n}, \mathsf{H}\right]
\end{aligned}
\end{equation}

\begin{equation} 
\begin{aligned}
\label{complete_version_constraint_per_device_battery}
 0 \leq E^B_{k,n} \leq B_{k}^{max} , \forall k \in \mathcal{K}     
 \end{aligned}
\end{equation}

where:
\begin{equation} \label{computation_time}
\begin{aligned}
 \tau_{k,n}= \frac{I_{k,n} \cdot \alpha \cdot s_{k,n}}{c_{k,n} \cdot f_{k,n}}, 
 \end{aligned}
\end{equation}

\begin{equation} \label{transmission_time}
\begin{aligned}
 tr_{k,n}= \frac{m}{r_{k,n}},
 \end{aligned}
\end{equation}

\begin{equation}
\begin{aligned}
 r_{k,n} = {b_{k,n} \cdot log_2\left(1 + \frac{g_{k,n} \cdot p_{k,n}}{b_{k,n} \cdot N_0}\right)}.
 \end{aligned}
\end{equation}

Constraint (\ref{complete_version_constraint_total_latency}) ensures the synchronization of the FL process by upper bounding the total time required by each worker to complete a computation and transmission task along with any queuing delay at the pre-selected time threshold $\mathsf{H}$.
Constraints (\ref{complete_version_constraint_per_device_capacity}) and (\ref{complete_version_constraint_per_device_power}) ensure that the computational capacity along with the transmission power of each worker $k$ at the $n^{th}$ \textit{global} iteration, are within the maximum available computation and communication capabilities, denoted by $f_{k,n}^{max}$ and $p_{k,n}^{max}$, respectively. Constraint (\ref{complete_version_constraint_idle}) ensures that at least one worker should be involved in the FL process. Constraint (\ref{complete_version_constraint_total_bandwidth}) ensures that the requested bandwidth of worker $k$, trying to access the channel, at each time $q$ ($q \in [\tau_{k,n}, \mathsf{H}]$) does not exceed the available bandwidth of the channel (right part of Eq. (\ref{complete_version_constraint_total_bandwidth})). The available bandwidth results from the maximum capacity of the channel at the $n^{th}$ \textit{global} iteration, notated as $b^{\text{max}}_n$ and the already allocated bandwidth from the rest of the workers that are already transmitting their model updates at time $q$. The $ch^{(q)}_{k,n}$ of constraint (\ref{complete_version_constraint_total_bandwidth}) is a Boolean indicating whether worker $k$ actively transmits its model updates at time $q$ at the $n^{th}$ \textit{global} iteration. Constraint (\ref{complete_version_constraint_per_device_battery}) ensures that the energy stored in the battery of worker $k$ at the $n^{th}$ global iteration $E^B_{k,n}$ is within the maximum battery capacity, denoted as $B^{\max}_k$.  

\vspace{-2pt}
\section{Proposed Deep Reinforcement Learning Solution}\label{proposed_solution}
\vspace{-3pt}
Reinforcement Learning (RL) is an exploration process aiming at maximizing a long-term reward through a sequence of interactions with an environment. On each step of interaction $n$ the agent observes a state $S_{n} \in \mathcal{S}$ that is the current representation of the environment, and selects an action $A_{n} \in \mathcal{A}$. At the next step $n+1$, as a consequence of the selected action, the agent receives a reward $\mathsf{r}_{n+1} \in \mathcal{R}$ (i.e. a numerical feedback) and transitions to the next state of the environment $S_{n+1} \in \mathcal{S}$ \cite{10.5555/3312046}. This sequential interaction underpins the agentic behavior of GreenFLag, enabling autonomous, reward-driven resource orchestration over successive FL iterations. We propose a Soft-Actor Critic (SAC) Deep Reinforcement Learning (DRL) solution to solve the optimization problem introduced in Section \ref{problem_formulation}, motivated by the fact that SAC is known to achieve efficient learning, stability and robustness \cite{haarnoja18b}. Our problem is episodic, with each episode ending in a terminal state defined by FL model convergence. The key terms are:\\
\textbf{Environment}: The coordinator and its workers participating in the FL process.\\
\textbf{Step}: One global FL iteration during which the RL agent observes the environment and allocates resources to workers.\\
\textbf{Episode}: A full FL process consisting of successive RL steps, terminating when the model reaches the target performance $\epsilon_0$.\\
\textbf{State:}  The information of the environment that the RL agent monitors at each RL step. It includes: the number of \textit{local} iterations taken place at each worker ($\mathrm{I}_{k,n-1}$), the amount of wasted energy consumed by each worker ($\mathrm{E}^W_{k,n-1}$), the global performance rate of the FL model ($\mathrm{e}_{n-1}$), the maximum available computation and communication capabilities of each worker ($\mathrm{f}_{k,n-1}^{max}$, $\mathrm{p}_{k,n-1}^{max}$), the size of each worker's local dataset ($|\mathcal{D}_{k,n-1}|$), the coordinator's maximum available bandwidth ($\mathrm{b}^{\text{max}}_{n-1}$), the available renewable energy ($\textbf{E}^R_{n-1}$), the maximum battery capacity ($\textbf{B}^{\text{max}}_{k}$) and the available energy stored in the battery ($\textbf{E}_{k,n-1}^B$) that the worker $k$ has access to.  The index $n-1$ denotes that these values correspond to the last \textit{global} iteration and are used by the agent as the observed state for selecting the action at iteration $n$. 

As a result, the state of the environment at the $n^{th}$ 
RL step is defined as:
\begin{equation*}
\resizebox{\columnwidth}{!}{$
 \mathcal{S}_n = \left\{
 \textbf{I}_{n-1}, \textbf{E}^W_{n-1}, \mathrm{e}_{n-1}, 
 \textbf{f}_{n-1}^{max}, \textbf{p}_{n-1}^{max},
 \textbf{D}_{n-1}, \mathrm{b}^{\text{max}}_{n-1},
 \textbf{E}^R_{n-1},\textbf{B}^{\text{max}}_{k}, \textbf{ E}^{B}_{k,n-1}
 \right\}
$}
\end{equation*}

where: \\ 
$\textbf{I}_{n-1} = [I_{1,n-1},...,I_{\mathcal{K},n-1}]^T$, 
$\textbf{E}^W_{n-1} = [E^W_{1,n-1},...,E^W_{\mathcal{K},n-1}]^T$, 
$\textbf{f}_{n-1}^{max} = [f^{max}_{1,n-1},...,f^{max}_{\mathcal{K},n-1}]^T$, 
$\textbf{p}_{n-1}^{max} = [p^{max}_{1,n-1},...,p^{max}_{\mathcal{K},n-1}]^T$,\\
$\textbf{D}_{n-1} = [|\mathcal{D}_{1,n-1}|,...,|\mathcal{D}_{\mathcal{K},n-1}|]^T$, \\
$\textbf{E}^R_{n-1} = [E^R_{1,n-1},...,E^R_{\mathcal{K},n-1}]^T$, 
$\textbf{B}^{\max}_{k} = [B^{\max}_1,...,B^{\max}_{\mathcal{K}}]^T$,\\
$\textbf{E}^{B}_{k,n-1} = [E^B_{1,n-1},...,E^B_{\mathcal{K},n-1}]^T$.\\\\
\textbf{Action Space:} The action space $\mathcal{A}_{n}$ is comprised of all computation and communication control actions that the GreenFLag agent will select for each worker at step $n$. Hence, the action space is formulated as: 
\begin{equation*}
 \mathcal{A}_{n} = \{\textbf{f}_n, \textbf{p}_n, \textbf{b}_n\}
\end{equation*}\\
Based on constraints (\ref{complete_version_constraint_per_device_capacity}), (\ref{complete_version_constraint_per_device_power}) and (\ref{complete_version_constraint_total_bandwidth}), the action space is bounded.\\
\textbf{Reward function:} The reward function of the RL agent is formulated based on the objective function Eq. (\ref{objective_function}), in conjunction with constraints (\ref{complete_version_constraint_total_latency}), (\ref{complete_version_constraint_idle}) and (\ref{complete_version_constraint_total_bandwidth}) (constraints (\ref{complete_version_constraint_per_device_capacity}) and (\ref{complete_version_constraint_per_device_power}) are satisfied by the bounded action space).
Specifically, the reward that the RL agent receives at the $n^{th}$ \textit{global} iteration is defined as:
\begin{equation*} \label{sac_reward}
\resizebox{\columnwidth}{!}{$
\mathsf{r}_n = - \left[ 
\displaystyle \sum_{k=1}^{\mathcal{K}} \Omega_{k,n} \cdot
\big( 
  E_{k,n}^C + E_{k,n}^T - (E_{k,n}^R + E_{k,n}^B) \big)_+ + x_n \right]
$}
\end{equation*}

In this work, the reward function encompasses both the total energy consumption of all workers participating in the FL process at step $n$ and the renewable energy. Additionally, $x_{n}$ is the penalty term defined to guarantee a safe RL process, ensuring that constraints (\ref{complete_version_constraint_total_latency}), (\ref{complete_version_constraint_idle}) and (\ref{complete_version_constraint_total_bandwidth}) are taken into account. As such, the penalty is defined as follows:
\begin{equation*} \label{penalty}
\begin{aligned}
  x_n = \sum_{k=1}^{\mathcal{K}} \left(E_{k,n}^W + \mu_1 \cdot P^{(1)}_{k,n} + \mu_3 \cdot P^{(3)}_{k,n,q} \cdot P^{(1)}_{k,n} \right) + \mu_2 \cdot P^{(2)}_n
  \end{aligned}
\end{equation*}

where $E_{k,n}^W$ is the amount of wasted computation energy in case the worker $k$ did not meet the time threshold at the $n^{th}$ RL step. Furthermore, $\mu_1$, $\mu_2$ and $\mu_3$ are constant penalty weights of each constraint violation. $P^{(1)}_{k,n}$, $P^{(2)}_{n}$ and $P^{(3)}_{k,n,q}$ are three indicator functions, related to constraints (\ref{complete_version_constraint_total_latency}), (\ref{complete_version_constraint_idle}) and (\ref{complete_version_constraint_total_bandwidth}). The $P^{(3)}_{k,n,q} \cdot P^{(1)}_{k,n}$ part of the equation assigns penalty to each worker $k$ that was not able to access the communication channel on time. 

\begin{equation} \label{indicator_function_time_violation}
  P_{k,n}^{(1)} =
  \left\{
  \begin{array}{ll}
    1, & \tau_{k,n} + tq_{k,n} + tr_{k,n} - \mathsf{H} \geq 0 \\
    0, & \text{otherwise}
  \end{array}
  \right.
\end{equation}

\begin{equation} \label{indicator_function_empty_selection}
  P_{n}^{(2)} =
  \left\{
  \begin{array}{ll}
    1, & \sum_{k=1}^{\mathcal{K}} f_{k,n} = 0 \\
    0, & \text{otherwise}
  \end{array}
  \right.
\end{equation}

\begin{equation}
\resizebox{\columnwidth}{!}{$
P^{(3)}_{k,n,q} =
\begin{cases}
1, & 
b_{k,n} - \Big( b^{\text{max}}_n - \sum_{i \in \mathcal{K}, i\neq k} \left( b_{i,n} \cdot ch^{(q)}_{k,n} \right)\Big) > 0, \ \forall q \in [\tau_{k,n},\mathsf{H}]\\[2pt]
0, & \text{otherwise}.
\end{cases}
$}
\end{equation}

Overall, the complete reward function is defined below:

\begin{equation} \label{sac_reward_expanded}
\resizebox{0.8\columnwidth}{!}{$
\begin{aligned}
r_n = - \Bigg[ 
& \sum_{k=1}^{\mathcal{K}} \Omega_{k,n} \cdot
\Big( E_{k,n}^C  + E_{k,n}^T - (E_{k,n}^R + E_{k,n}^B) \Big)_+ + \\
& + E_{k,n}^W 
+ (\mu_1 + \mu_3 \cdot P^{(3)}_{k,n,q}) \cdot P^{(1)}_{k,n} 
\Big] 
+ \mu_2 \cdot P^{(2)}_n 
\Bigg]
\end{aligned}
$}
\end{equation}

\section{Simulation Setup} \label{sim_setup}
The current section provides the simulation setup used to evaluate the performance of the proposed safe RL solution.  

\textbf{Network Environment Setup:}
The wireless communication environment consists of one coordinator and 20 heterogeneous workers ($\mathcal{K}=20$). Up to 60\% of the workers are low-end, and in each experiment their exact number is drawn from a truncated normal distribution
\cite{Robert1995}. 
Each low-end worker $i$ has a maximum available computation and communication capacity which is uniformly selected in the range $f_{i,n}^{max} \in [1,3]$ GHz and $p_{i,n}^{max} \in [23,28]$ dBm, with a total number of FLOPs per cycle equal to $C_i = 4$. 
In the same notion, the resources of the high-end devices $j$ are selected in the range $f_{j,n}^{max} \in [3.2,5]$ GHz, and $p_{j,n}^{max} \in [29,33]$ dBm, respectively, with a total number of FLOPs per cycle equal to $C_j = 2$. 
The effective switched capacitance is fixed for all workers and equal to $\varsigma_k = 10^{-28}$ $\mathrm{Watt/Hz^{3}}$ \cite{Yang13037554}.
The channel gain is modeled as $g_{k,n} = 127 + 30log_{10}(d_{k,n})$, where $N_0= -158$ dBm/Hz is the white Gaussian noise power spectral density \cite{9462445Zhou} and $d_{k,n}$ is the distance of worker $k$ from the coordinator. 
The distance $d_{k,n}$ fluctuates in the range $[10,500]$ meters, forming an environment where mobile workers exist. 
The maximum available bandwidth $b^{\max}_n$ at each global iteration spans in the range $[50,100] $ MHz. 

\textbf{Renewable Sources Setup:}
The Copernicus dataset is utilized to collect the necessary atmospheric related data towards simulating the renewable energy production from solar panels and wind turbines. The selected dataset consists of hourly sensor measurements in the area of Athens, Attica for the year 2020 \cite{copernicus}. \\
\underline{Solar Energy}: 
To simulate and calculate the solar energy production $E^{S}_{k,n}$ in (\ref{solar_energy}), the direct solar radiation $P^{SA}_{n}$ and the total cloud coverage $N_n$ of each global iteration are retrieved from the Copernicus dataset. The solar panel area is set to $\mathsf{z} = 0.03 m^2$.
\\
\underline{Wind Energy}:
The wind energy production $E^{WN}_{k,n}$ as shown in (\ref{wind}), is simulated considering the total wind speed $\mathsf{v}$ from the Copernicus dataset. The configuration parameters for the air density and the blade swept area are set to $\rho = 1.225 \frac{kg}{m^3}$ and $S = 0.1 m^2$, respectively. 

Based on the above configurations, the total renewable energy production $E^R_{k,n} $ is computed from the Copernicus dataset,  over each time period $\mathsf{H}$. The measurements from the dataset are highly connected to the day, time and month of collection. In practice, edge devices cannot utilize the full amount of renewable energy made available at their location. This could occur due to hardware restrictions, limited harvesting efficiency, and power-conversion constraints. To reflect these realistic device-level limitations, we consider scenarios in Section \ref{performance_eval} in which the renewable resources are sparse and limited. Each worker is connected to a unique, local battery with stored energy $E^B_{k,n}$ and maximum capacity $B^{\max}_k \in [15,50] $ J. Each worker begins the RL episode with different battery levels $E^B_{k,n}$, proportionally set to their maximum capacities $B^{\max}_k$. To emulate occasional hardware or communication failures among workers and their energy sources, in each RL episode a sequence of global iterations is randomly selected during which the harvested renewable energy is temporarily unavailable ($E^R_{k,n} = 0$).

\textbf{Federated Learning Setup:}
The FL process considers a Convolutional Neural Network (CNN) for handwritten digit recognition, using the MNIST dataset \cite{lecun2010mnist}. Table \ref{fl_model} highlights the FL configuration for our setup.

\begin{table}[ht!]
\renewcommand{\arraystretch}{1.05}
\centering
\begin{tabular}{|c|c|} 
\hline
\textbf{Architecture} & \textbf{Configuration} \\ [0.7ex]
\hline
Total / Per Worker Samples & 60,000 / [200,800] \\
\hline
Trainable Parameters / Model Size & 658,922 / 2.51 MB\\
\hline
Model Complexity &  1.8 MFLOPs\\
\hline
Batch Size / Optimizer / Learning Rate& 32 / Adam / $5 \cdot 10^{-5}$ \\
\hline
Activation Function & ReLU \\
\hline
$\eta$ / $\epsilon_0$ / $\mathsf{H}$ & 0.5 / 0.04 / 20 sec \\
\hline

\end{tabular}
\caption{FL Configuration}
\label{fl_model}
\vspace{-3pt}
\end{table}


\textbf{Reinforcement Learning Setup:} 
The proposed DRL solution is based on a custom environment interfaced through Stable-Baselines3\cite{Raffin3546526}, leveraging a Soft-Actor Critic RL approach. 
The RL agent is trained offline, interacting with an environment that follows the system model and problem formulation from Sections \ref{system_model} and \ref{problem_formulation}. For the training phase of the RL agent, a simulated FL environment was used, as proposed in \cite{koursioumpas2024safe}. The adaptation of this approach significantly benefits the training process, as it avoids the overhead of executing real FL processes. In the meantime, the RL agent is exposed to a wide range of heterogeneous system conditions (e.g. device capabilities, channel conditions and renewable energy availability). As a result, the RL agent experiences a broad spectrum of system configurations, which improves policy robustness and generalization to unseen conditions.  
The learned policy is then used in all experiments in Section \ref{performance_eval} without any further fine-tuning.

The hyperparameters \cite{sac_baselines3} of the DRL algorithm have been tuned, based on experimentation and are summarized in Table \ref{sac_model}. 
The agent architecture comprises one Policy, two Value and three Target Multi-Layer Perceptron (MLP) Networks, and optimizes a stochastic policy. 
The optimizer selected for the training phase is the Adam with initial value of learning rate set to 0.001 and the batch size is set to 256. 
A step decay scheduler is used to reduce the learning rate every 6000 RL episodes by 1\%. 
The entropy regularization coefficient that controls the exploration/exploitation trade-off of the RL agent is set to 0.8. 
The RL agent is trained every 1000 RL steps and the training starts after the first 100 RL steps, in order to fill the replay buffer of size $2\cdot10^{6}$ with enough samples. 
The weight hyperparameters \{$\mu_1$, $\mu_2$, $\mu_3$\} of the penalty function have been set to  \{0.3, 0.4, 0.3\}.

\begin{table}[ht!]

\renewcommand{\arraystretch}{1.05}
\centering
\begin{tabular}{|c|c|} 
\hline
\textbf{Architecture} & \textbf{Configuration} \\ [0.7ex]
\hline
 Model / Policy & MLP / Stochastic\\
\hline
Policy / Value / Target DNNs & 1 / 2 / 3\\
\hline
Hidden layers / Neurons &  8 / 512\\
\hline
Batch Size / Optimizer/ Learning Rate& 256 / Adam / 0.001 \\
\hline
Learning Rate Scheduler: Step Decay / Drop rate & 6000 episodes / 1$\%$  \\
\hline
Entropy Coefficient & auto\_0.8 \\
\hline
Training / SDE Sample Frequency & 1000 / 100 RL steps \\
\hline
Training Starts & 100 RL steps \\
\hline
Replay Buffer Size & $2\cdot10^{6}$ \\
\hline
\end{tabular}
\caption{RL Agent Configuration}
\label{sac_model}
\vspace{-7pt}
\end{table}

\section{Performance Evaluation} \label{performance_eval}
To evaluate the reduction in total grid energy consumption achieved by GreenFLag, we define the following scenarios:
\begin{enumerate}
    \item Scenario 1: 
    This scenario acts as a best case for the evaluation, presenting ideal conditions in terms of renewable availability, as it considers that all workers start with a fully charged battery and all can harvest during an FL process execution. 
    \item Scenario 2: This scenario is selected to evaluate the adaptability of our approach in more realistic conditions, where the availability of renewables is more limited (e.g. unfavorable weather conditions, hardware malfunctions, limited stored energy). This is realized by introducing sporadic renewable outages for up to $60\%$ of the workers, while the rest can harvest renewable energy. The affected subset of workers begin the FL process with fully charged batteries, while the rest begin with a random battery percentage of $[50,100]\%$. 
    \item Scenario 3: This scenario acts as the worst-case one and evaluates the robustness of our solution, when part of the system is fully grid-dependent. More specifically, a subset of workers, up to $60\%$, has no renewable energy available (zero harvested and stored energy) throughout an FL process, while the rest can harvest and have a battery percentage of $[50,100]\%$. 
\end{enumerate}
Then, a comparison between GreenFLag and three baselines schemes is performed. The comparison is with regard to \textbf{1)} total energy consumed in Joules, \textbf{2)} the energy consumed from the grid, \textbf{3)} the energy consumed from renewable resources, \textbf{4)} the total duration of the global iteration in seconds, \textbf{5)} violations averaged per worker, and \textbf{6)} number of global iterations. All statistical results are averaged, including also their standard deviation ($\pm$STD) over 100 independent FL executions, in order to draw accurate conclusions.
We compare GreenFLag against a Best Effort, a Random Selection and a Greedy Selection Scheme.  \cite{9384231Zhan}. 
\begin{itemize}
\item \textbf{Best Effort Scheme (BES): } The BES selects in each \textit{global} iteration of an FL process, the maximum available capacities of each worker, without considering the energy aspect of the system. The objective of such scheduler is the acceleration of the FL process.
\item \textbf{Random Selection Scheme (RSS): } In each \textit{global} iteration, the RSS orchestrates randomly the resources of each worker, based on their available capacities. 
\item \textbf{Greedy Selection Scheme (GSS): } In each \textit{global} iteration, the GSS chooses the resource capacities of all workers that led to the best outcome so far, in terms of the total energy consumption. 
\end{itemize}

Tables \ref{scen1-comp}, \ref{scen2-comp}, \ref{scen3-comp} showcase the performance evaluation of GreenFLag (GFL) against the three baseline schemes under the considered scenarios. As it can be deduced from the tables, GreenFLag outperforms all three baselines in terms of both grid and total energy consumption. 

\begin{table}[ht!]
\begin{tabular}{|l|cccc|}
\hline
\textbf{\begin{tabular}[c]{@{}l@{}}Total Avg.\\ ($\pm$ STD)\end{tabular}} &
  \multicolumn{4}{c|}{\textbf{Scenario 1}} \\ \hline
\textbf{} &
  \multicolumn{1}{c|}{\cellcolor[HTML]{EFEFEF}\textbf{GFL}} &
  \multicolumn{1}{c|}{\textbf{BES}} &
  \multicolumn{1}{c|}{\textbf{RSS}} &
  \textbf{GSS} \\ \hline
\textbf{\begin{tabular}[c]{@{}l@{}}Total \\ Energy (J)\end{tabular}} &
  \multicolumn{1}{c|}{\cellcolor[HTML]{EFEFEF}\begin{tabular}[c]{@{}c@{}}258.1\\ ($\pm$ 110.7)\end{tabular}} &
  \multicolumn{1}{c|}{\begin{tabular}[c]{@{}c@{}}3801.4\\ ($\pm$ 848)\end{tabular}} &
  \multicolumn{1}{c|}{\begin{tabular}[c]{@{}c@{}}556.6\\ ($\pm$ 91.7)\end{tabular}} &
  \begin{tabular}[c]{@{}c@{}}382.8\\ ($\pm$ 93.8)\end{tabular} \\ \hline
\textbf{\begin{tabular}[c]{@{}l@{}}Grid \\ Energy (J)\end{tabular}} &
  \multicolumn{1}{c|}{\cellcolor[HTML]{EFEFEF}\begin{tabular}[c]{@{}c@{}}3.7\\ ($\pm$ 6.3)\end{tabular}} &
  \multicolumn{1}{c|}{\begin{tabular}[c]{@{}c@{}}3054.9\\ ($\pm$ 884.7)\end{tabular}} &
  \multicolumn{1}{c|}{\begin{tabular}[c]{@{}c@{}}107.7\\ ($\pm$ 45)\end{tabular}} &
  \begin{tabular}[c]{@{}c@{}}30\\ ($\pm$ 7.8)\end{tabular} \\ \hline
\textbf{\begin{tabular}[c]{@{}l@{}}Green \\ Energy (J)\end{tabular}} &
  \multicolumn{1}{c|}{\cellcolor[HTML]{EFEFEF}\begin{tabular}[c]{@{}c@{}}254.4\\ ($\pm$ 113)\end{tabular}} &
  \multicolumn{1}{c|}{\begin{tabular}[c]{@{}c@{}}746.4\\ ($\pm$ 108.9)\end{tabular}} &
  \multicolumn{1}{c|}{\begin{tabular}[c]{@{}c@{}}448.9\\ ($\pm$ 61.8)\end{tabular}} &
  \begin{tabular}[c]{@{}c@{}}352.8\\ ($\pm$ 93.4)\end{tabular} \\ \hline
\textbf{\begin{tabular}[c]{@{}l@{}}Duration of\\ Global\\ Iteration (s)\end{tabular}} &
  \multicolumn{1}{c|}{\cellcolor[HTML]{EFEFEF}\begin{tabular}[c]{@{}c@{}}13.8\\ ($\pm$ 0.8)\end{tabular}} &
  \multicolumn{1}{c|}{\begin{tabular}[c]{@{}c@{}}16.1\\ ($\pm$ 2.9)\end{tabular}} &
  \multicolumn{1}{c|}{\begin{tabular}[c]{@{}c@{}}13.3\\ ($\pm$ 2.1)\end{tabular}} &
  \begin{tabular}[c]{@{}c@{}}13.2\\ ($\pm$ 1.5)\end{tabular} \\ \hline
\textbf{\begin{tabular}[c]{@{}l@{}}Violations per\\ Worker\end{tabular}} &
  \multicolumn{1}{c|}{\cellcolor[HTML]{EFEFEF}\begin{tabular}[c]{@{}c@{}}0.4\\ ($\pm$ 0.2)\end{tabular}} &
  \multicolumn{1}{c|}{\begin{tabular}[c]{@{}c@{}}0\\ ($\pm$ 0)\end{tabular}} &
  \multicolumn{1}{c|}{\begin{tabular}[c]{@{}c@{}}1\\ ($\pm$ 0.4)\end{tabular}} &
  \begin{tabular}[c]{@{}c@{}}0.4\\ ($\pm$ 0.2)\end{tabular} \\ \hline
\textbf{\begin{tabular}[c]{@{}l@{}}Global \\ Iterations\end{tabular}} &
  \multicolumn{1}{c|}{\cellcolor[HTML]{EFEFEF}\begin{tabular}[c]{@{}c@{}}12.3\\ ($\pm$ 3.3)\end{tabular}} &
  \multicolumn{1}{c|}{\begin{tabular}[c]{@{}c@{}}11.6\\ ($\pm$ 1.2)\end{tabular}} &
  \multicolumn{1}{c|}{\begin{tabular}[c]{@{}c@{}}11.4\\ ($\pm$ 0.9)\end{tabular}} &
  \begin{tabular}[c]{@{}c@{}}11.4\\ ($\pm$ 0.8)\end{tabular} \\ \hline
\end{tabular}
\caption{\textbf{Scenario 1}: Comparison Results}
\label{scen1-comp}
\vspace{-5pt}
\end{table}

\begin{table}[ht!]
\begin{tabular}{|l|cccc|}
\hline
\textbf{\begin{tabular}[c]{@{}l@{}}Total Avg.\\ ($\pm$ STD)\end{tabular}} &
  \multicolumn{4}{c|}{\textbf{Scenario 2}} \\ \hline
\textbf{} &
  \multicolumn{1}{c|}{\cellcolor[HTML]{EFEFEF}\textbf{GFL}} &
  \multicolumn{1}{c|}{\textbf{BES}} &
  \multicolumn{1}{c|}{\textbf{RSS}} &
  \textbf{GSS} \\ \hline
\textbf{\begin{tabular}[c]{@{}l@{}}Total \\ Energy (J)\end{tabular}} &
  \multicolumn{1}{c|}{\cellcolor[HTML]{EFEFEF}\begin{tabular}[c]{@{}c@{}}240.7\\ ($\pm$ 48.1)\end{tabular}} &
  \multicolumn{1}{c|}{\begin{tabular}[c]{@{}c@{}}3774.1\\ ($\pm$ 390.6)\end{tabular}} &
  \multicolumn{1}{c|}{\begin{tabular}[c]{@{}c@{}}617.2\\ ($\pm$ 107.5)\end{tabular}} &
  \begin{tabular}[c]{@{}c@{}}572.2\\ ($\pm$ 66.9)\end{tabular} \\ \hline
\textbf{\begin{tabular}[c]{@{}l@{}}Grid \\ Energy (J)\end{tabular}} &
  \multicolumn{1}{c|}{\cellcolor[HTML]{EFEFEF}\begin{tabular}[c]{@{}c@{}}11.7\\ ($\pm$ 15)\end{tabular}} &
  \multicolumn{1}{c|}{\begin{tabular}[c]{@{}c@{}}3156.4\\ ($\pm$ 383.9)\end{tabular}} &
  \multicolumn{1}{c|}{\begin{tabular}[c]{@{}c@{}}156.1\\ ($\pm$ 110.2)\end{tabular}} &
  \begin{tabular}[c]{@{}c@{}}192.3\\ ($\pm$ 80.8)\end{tabular} \\ \hline
\textbf{\begin{tabular}[c]{@{}l@{}}Green \\ Energy (J)\end{tabular}} &
  \multicolumn{1}{c|}{\cellcolor[HTML]{EFEFEF}\begin{tabular}[c]{@{}c@{}}229\\ ($\pm$ 35.4)\end{tabular}} &
  \multicolumn{1}{c|}{\begin{tabular}[c]{@{}c@{}}617.6\\ ($\pm$ 129.5)\end{tabular}} &
  \multicolumn{1}{c|}{\begin{tabular}[c]{@{}c@{}}461.2\\ ($\pm$ 32.9)\end{tabular}} &
  \begin{tabular}[c]{@{}c@{}}379.9\\ ($\pm$ 40.4)\end{tabular} \\ \hline
\textbf{\begin{tabular}[c]{@{}l@{}}Duration of \\ Global \\Iteration (s)\end{tabular}} &
  \multicolumn{1}{c|}{\cellcolor[HTML]{EFEFEF}\begin{tabular}[c]{@{}c@{}}14\\ ($\pm$ 0.8)\end{tabular}} &
  \multicolumn{1}{c|}{\begin{tabular}[c]{@{}c@{}}16.9\\ ($\pm$ 2.4)\end{tabular}} &
  \multicolumn{1}{c|}{\begin{tabular}[c]{@{}c@{}}13.4\\ ($\pm$ 1.5)\end{tabular}} &
  \begin{tabular}[c]{@{}c@{}}14.8\\ ($\pm$ 1.3)\end{tabular} \\ \hline
\textbf{\begin{tabular}[c]{@{}l@{}}Violations per\\ Worker\end{tabular}} &
  \multicolumn{1}{c|}{\cellcolor[HTML]{EFEFEF}\begin{tabular}[c]{@{}c@{}}0.3\\ ($\pm$ 0.1)\end{tabular}} &
  \multicolumn{1}{c|}{\begin{tabular}[c]{@{}c@{}}0\\ ($\pm$ 0)\end{tabular}} &
  \multicolumn{1}{c|}{\begin{tabular}[c]{@{}c@{}}0.9\\ ($\pm$ 3)\end{tabular}} &
  \begin{tabular}[c]{@{}c@{}}0.3\\ ($\pm$ 0.2)\end{tabular} \\ \hline
\textbf{\begin{tabular}[c]{@{}l@{}}Global \\ Iterations\end{tabular}} &
  \multicolumn{1}{c|}{\cellcolor[HTML]{EFEFEF}\begin{tabular}[c]{@{}c@{}}10.6\\ ($\pm$ 0.8)\end{tabular}} &
  \multicolumn{1}{c|}{\begin{tabular}[c]{@{}c@{}}10.7\\ ($\pm$ 0.6)\end{tabular}} &
  \multicolumn{1}{c|}{\begin{tabular}[c]{@{}c@{}}11.4\\ ($\pm$ 1.4)\end{tabular}} &
  \begin{tabular}[c]{@{}c@{}}12.6\\ ($\pm$ 1)\end{tabular} \\ \hline
\end{tabular}
\caption{\textbf{Scenario 2}: Comparison Results}
\label{scen2-comp}
\vspace{-5pt}
\end{table}

\begin{table}[ht!]
\begin{tabular}{|l|cccc|}
\hline
\textbf{\begin{tabular}[c]{@{}l@{}}Total Avg.\\ ($\pm$ STD)\end{tabular}} &
  \multicolumn{4}{c|}{\textbf{Scenario 3}} \\ \hline
\textbf{} &
  \multicolumn{1}{c|}{\cellcolor[HTML]{EFEFEF}\textbf{GFL}} &
  \multicolumn{1}{c|}{\textbf{BES}} &
  \multicolumn{1}{c|}{\textbf{RSS}} &
  \textbf{GSS} \\ \hline
\textbf{\begin{tabular}[c]{@{}l@{}}Total \\ Energy (J)\end{tabular}} &
  \multicolumn{1}{c|}{\cellcolor[HTML]{EFEFEF}\begin{tabular}[c]{@{}c@{}}205.8\\ ($\pm$ 45.2)\end{tabular}} &
  \multicolumn{1}{c|}{\begin{tabular}[c]{@{}c@{}}3375.7\\ ($\pm$ 837.4)\end{tabular}} &
  \multicolumn{1}{c|}{\begin{tabular}[c]{@{}c@{}}578.8\\ ($\pm$ 109.2)\end{tabular}} &
  \begin{tabular}[c]{@{}c@{}}474.9\\ ($\pm$ 96.4)\end{tabular} \\ \hline
\textbf{\begin{tabular}[c]{@{}l@{}}Grid \\ Energy (J)\end{tabular}} &
  \multicolumn{1}{c|}{\cellcolor[HTML]{EFEFEF}\begin{tabular}[c]{@{}c@{}}13.6\\ ($\pm$ 16.7)\end{tabular}} &
  \multicolumn{1}{c|}{\begin{tabular}[c]{@{}c@{}}2814.5\\ ($\pm$ 793.4)\end{tabular}} &
  \multicolumn{1}{c|}{\begin{tabular}[c]{@{}c@{}}173.5\\ ($\pm$ 104.2)\end{tabular}} &
  \begin{tabular}[c]{@{}c@{}}164.8\\ ($\pm$ 96.3)\end{tabular} \\ \hline
\textbf{\begin{tabular}[c]{@{}l@{}}Green \\ Energy (J)\end{tabular}} &
  \multicolumn{1}{c|}{\cellcolor[HTML]{EFEFEF}\begin{tabular}[c]{@{}c@{}}192.3\\ ($\pm$ 42)\end{tabular}} &
  \multicolumn{1}{c|}{\begin{tabular}[c]{@{}c@{}}561.2\\ ($\pm$ 119.1)\end{tabular}} &
  \multicolumn{1}{c|}{\begin{tabular}[c]{@{}c@{}}405.3\\ ($\pm$ 99)\end{tabular}} &
  \begin{tabular}[c]{@{}c@{}}310.1\\ ($\pm$ 50.1)\end{tabular} \\ \hline
\textbf{\begin{tabular}[c]{@{}l@{}}Duration of \\ Global\\Iteration (s)\end{tabular}} &
  \multicolumn{1}{c|}{\cellcolor[HTML]{EFEFEF}\begin{tabular}[c]{@{}c@{}}14.3\\ ($\pm$ 0.6)\end{tabular}} &
  \multicolumn{1}{c|}{\begin{tabular}[c]{@{}c@{}}16.6\\ ($\pm$ 2.5)\end{tabular}} &
  \multicolumn{1}{c|}{\begin{tabular}[c]{@{}c@{}}14.1\\ ($\pm$ 1.5)\end{tabular}} &
  \begin{tabular}[c]{@{}c@{}}13.1\\ ($\pm$ 1.9)\end{tabular} \\ \hline
\textbf{\begin{tabular}[c]{@{}l@{}}Violations per\\ Worker\end{tabular}} &
  \multicolumn{1}{c|}{\cellcolor[HTML]{EFEFEF}\begin{tabular}[c]{@{}c@{}}0.4\\ ($\pm$ 0.1)\end{tabular}} &
  \multicolumn{1}{c|}{\begin{tabular}[c]{@{}c@{}}0\\ ($\pm$ 0)\end{tabular}} &
  \multicolumn{1}{c|}{\begin{tabular}[c]{@{}c@{}}0.9\\ ($\pm$ 3)\end{tabular}} &
  \begin{tabular}[c]{@{}c@{}}0.2\\ ($\pm$ 0.2)\end{tabular} \\ \hline
\textbf{\begin{tabular}[c]{@{}l@{}}Global \\ Iterations\end{tabular}} &
  \multicolumn{1}{c|}{\cellcolor[HTML]{EFEFEF}\begin{tabular}[c]{@{}c@{}}10.5\\ ($\pm$ 0.9)\end{tabular}} &
  \multicolumn{1}{c|}{\begin{tabular}[c]{@{}c@{}}10.2\\ ($\pm$ 0.9)\end{tabular}} &
  \multicolumn{1}{c|}{\begin{tabular}[c]{@{}c@{}}11.2\\ ($\pm$ 1.2)\end{tabular}} &
  \begin{tabular}[c]{@{}c@{}}11.8\\ ($\pm$ 0.8)\end{tabular} \\ \hline
\end{tabular}
\caption{\textbf{Scenario 3}: Comparison Results}
\label{scen3-comp}
\end{table}

\textbf{Key Observation 1}:
In all scenarios, GreenFLag consistently demonstrates substantial energy efficiency gains relative to the baseline schemes. As scenarios become progressively more demanding, renewable availability decreases. Consequently, the system relies on grid power, leading to higher grid energy consumption. However, GreenFLag exhibits only a small increase compared to the baselines, indicating strong robustness to renewable scarcity. 

\textbf{Key Observation 2}: GreenFLag achieves lower total energy consumption in Scenarios 2 and 3 compared to Scenario 1, even though renewable availability is reduced and grid energy usage is higher. This behavior results from the reward function, which explicitly penalizes grid energy consumption. Whenever grid energy is utilized, GreenFLag is incentivized to adapt its decisions to reduce grid usage, which in turn drives down the overall energy consumption. This reduction causes a slight increase in the global iteration duration yet it remains comparable to the baseline solutions. 

\textbf{Key Observation 3}: Although BES exploits the maximum capabilities of each worker, it attempts to serve as many workers as possible simultaneously by allocating lower bandwidth. This results in increased transmission time, making it the slowest scheduler. The same strategy causes excessive reliance on grid energy, explaining its substantially higher grid energy consumption, despite achieving zero violations.

\textbf{Key Observation 4}: Despite the fact that GSS performs best, its decisions are based solely on immediate per-round energy outcomes. As a result, it fails to capture the long-term impact of these decisions on cumulative energy consumption and grid dependence over the FL process. GreenFLag overcomes these limitations by optimizing for a long-term objective, leading to significantly lower grid energy consumption than BES, RSS, and GSS.

\textbf{Key Observation 5}: GreenFLag reduces grid-energy consumption by approximately 8-16 times compared to the best-performing baseline (GSS), and simultaneously lowers the total system energy consumption by a factor of 1.5-2.4, depending on the scenario. 

\textbf{Key Observation 6}: GreenFLag exhibits nearly identical convergence behavior across all evaluated scenarios. The number of global iterations required to reach the target model performance is comparable to the baseline solutions. These results suggest that the energy-efficient strategy does not negatively impact the learning process of the FL and that energy savings are not achieved by slowing convergence or reducing model accuracy.

To sum up, GreenFLag consistently attains significantly lower grid and total energy than all three baselines, while preserving comparable convergence speed, global iteration duration and number of violations. 

\section{Conclusions}
This paper proves the effectiveness and feasibility of GreenFLag, a safe DRL-based resource orchestration agent for federated learning in wireless networks, with the objective of minimizing reliance on grid power by integrating renewable energy sources into the system and optimizing resource allocation toward carbon-efficient operation. A penalty function is introduced to ensure a safe RL process by enforcing feasibility with respect to system constraints and by discouraging wasted energy usage, while jointly minimizing the grid energy consumption of the system.
An FCFS scheduler is adopted to ensure realistic bandwidth allocation. The renewable resources are obtained by simulating realistic conditions using the Copernicus dataset. Overall, the evaluation results demonstrate the effectiveness of GreenFLag in reducing grid energy consumption through renewable-aware resource orchestration by $94.8\%$ on average, while maintaining feasible solutions with respect to system constraints. The learned policy exhibits low violation rates across all scenarios, confirming its applicability under realistic network conditions. More importantly, these energy and carbon-footprint savings are achieved without compromising FL accuracy or convergence speed. 

\section{Acknowledgments}
This work has been funded by the European Commission Horizon Europe Smart Networks and Services Joint Undertaking (SNS JU) EXIGENCE Project (Grant Agreement No. 101139120). 

\bibliographystyle{IEEEtran}
\bibliography{bibliography/bib}

\end{document}